\documentclass[a4paper,11pt]{article}
\usepackage{pos}
\def\Re{\text{Re}}
\DeclareMathOperator{\Tr}{Tr}

\title{Simple implementation of color
coherence for the resumation of
soft BDMPS-Z gluons}

\author*[a]{João Barata}
\author[a]{Fabio  Domínguez}
\author[a]{Carlos Salgado}
\author[a]{Víctor Vila}

\affiliation[a]{Instituto Galego de Física de Altas Enerxías (IGFAE), Universidade de Santiago de Compostela,
E-15782 Galicia, Spain}

\emailAdd{joaolourenco.henriques@usc.es}

\abstract{The evolution of QCD jets under the influence of a dense colored medium leads to the non trivial modification of the gluon emission spectrum. In the multiple soft scattering regime, for sufficiently large media, soft and wide angle emissions can be resummed (in the planar limit of QCD) due to the small formation time of each gluon emission. Similarly to DGLAP evolution, such cascades correspond to pure Markovian processes and interferences between partons are neglected. In this strict limit, color coherence effects are absent. In the vacuum this coherence effect is critical to guarantee that the shower is angular ordered. In the medium, it is known that such color coherence effects might play a role in suppressing radiation at intermediate emission angles.\par
We present a simple procedure to implement color (de)coherence effects into the in-medium evolution equations obtained in the pure Markovian approximation. We verify that the result obtained is in accordance with previous studies regarding the role of in-medium color coherence in the evolution of the QCD antenna. The new evolution equation incorporates a novel emission kernel which enforces color coherence between emitters.}

\FullConference{%
  HardProbes2020\\
  1-6 June 2020\\
  Austin, Texas}

\begin{document}
\maketitle

\section{Introduction}
When a hard probe propagates through a dense and (sufficiently) long medium the contribution to the gluon emission spectrum due to the multiple (soft) scatterings (in opposition to single rare hard scattering) becomes important and has to be taken into account. In the BDMPS-Z approximation one is able to resum such interactions by assuming that the scattering potential is harmonic. Parametrically, the gluon emission spectrum in this approximation takes the form \cite{BDIM1}
\begin{equation}
\omega\frac{\rm{d} I}{\rm{d} \omega} \sim \alpha_s\sqrt{\frac{\omega_c}{\omega}} \quad (\omega<\omega_c) \, ,
\end{equation}
where $\omega_c=\frac{1}{2}\hat{q}L^2$ is the frequency of gluons whose formation time $t_f$ is of the order of $L$, with $\hat{q}$ the medium diffusion coefficient and $L$ the medium length. The scale $\omega_c$ divides the spectrum into two regions: gluons with energy $\omega>\omega_c$ are highly suppressed and populate the collinear region ($\theta<\theta_c$, where $\theta$ is the emission angle and $\theta_c\sim \frac{1}{\hat{q}L^3}$ the emission angle for gluon of frequency $\omega_c$)  while the region $\omega\ll\omega_c$ is highly populated by large angle gluons ($\theta\gg\theta_c$).\par 
Since the typical formation time for in-medium gluon emissions satisfies $t_f\sim \sqrt{\frac{\omega}{\hat{q}}}$, these gluons are formed almost instantaneously (i.e. $\frac{t_f}{L}\ll1$). As a consequence, since (parametrically) the phase space for in-medium emission must scale as $L-t_f\sim L$, we have that each gluon is emitted decoherently from all the other emissions at any point in the medium. This is the physical motivation for the Poissonian/Markovian approximation to in-medium gluon emission.\par 
One of the major breakthroughs beyond the BDMPS-Z framework came from the study of the QCD antenna (i.e. a boson decaying to a quark dipole) in-medium followed by a soft gluon emission. Such a set up was previously consider in the context of vacuum QCD and lead to the discovery of angle ordering, which is a critical effect one must consider in order to describe data. In the case of in-medium evolution, the same set up lead to the following conclusion: if the emission angle is small compared to $\theta_c$ than the systems keeps color coherence and the new pair of particles emits as if they are a single color charge; in the opposite limit when the angle of emission is larger than $\theta_c$ the emitters lose color coherence and the spectrum of emission will be that of two independent color charges \cite{Antenna4}. The interference terms present in the spectrum of emission are thus proportional to
\begin{equation}\label{eq:deltam}
1-e^{-\left(\frac{\theta}{\theta_c}\right)^2}  \sim \Theta(\theta>\theta_c)  \, , 
\end{equation}
where the last term symbolizes the typical step function approximation described above.\par 
Combining these results with the above discussion regarding interferences due to the quantum mechanical formation time of radiation we see that the phase space for emission is now proportional to $L- \min(t_f,t_d)\sim L$ (where $t_d$ is the typical decoherence time obtained from \eqref{eq:deltam}).\par 
We go beyond the completely decoherent picture by including the corrections due to color coherence between emitters, while still resuming all the soft gluon emissions. To do this, as we detail below, we compute the BDMPS-Z spectrum followed by a soft gluon emission and try to simplify the color structure of the process such that a piece controlling coherence can be extracted. The phase space of emission is thus proportional to $L-t_d$. This result is then applied to the rate equations.\par 
\section{Gluon in-medium emission followed by a soft gluon}
\begin{figure}[h!]
    \centering
    \includegraphics[scale=0.9]{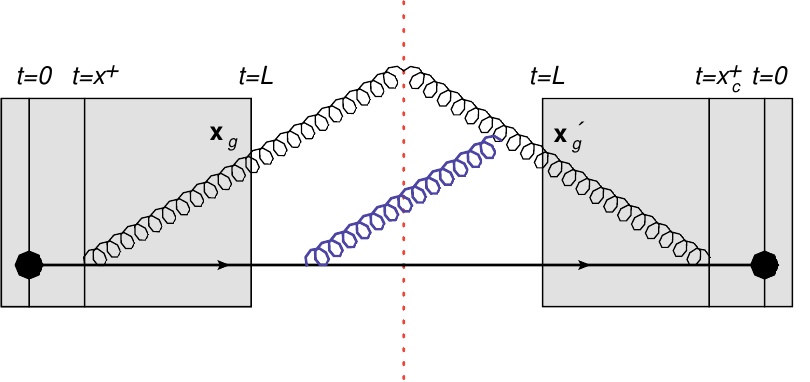}
    \caption{Interference diagram due to the presence of an extra soft vacuum gluon.}
    \label{fig:1}
\end{figure}
We explicitly compute the diagram present in figure \ref{fig:1} and obtain
\begin{equation}
 \begin{split}\label{eq:main_amplitude_2}
&\omega \omega^\prime \frac{dI}{d^2\mathbf{k}d^2\mathbf{k}^\prime d\omega d\omega^\prime}=\left(\omega^\prime \frac{dI}{d\omega^\prime d^2\mathbf{k}^{\prime}}\right)^{\mathbf{g}}\frac{2C_F\alpha_s}{(2\pi^2)\omega^{2}N_c(N_c^2-1)^2 }\Re\Bigg[\int_{\textbf{x}_g\textbf{x}_g^\prime x^+ x^{+}_c \textbf{z}}e^{i\textbf{k}(\textbf{x}^\prime_g-\textbf{x}_g)}
\\ &\times \partial_{\textbf{x}}\mid_0\cdot \partial_{\textbf{x}^\prime}\mid_0 \Tr \langle G(\textbf{x},\textbf{z}\mid \omega) W(\textbf{0}))\rangle_{x^+_c,x^+}  f^{jia}f^{cbd}    \langle W^{ac}(\textbf{0}) G^{jb}(\textbf{z},\textbf{x}_g\mid \omega)G^{\dagger id}(\textbf{x}^\prime_g,\textbf{x}^\prime\mid \omega) \rangle_{L,x_c^+}\Bigg] \, ,
\end{split}
\end{equation}
which should be directly compared to the BDMPS-Z equivalent diagram (no blue gluon in figure \ref{fig:1})
\begin{equation}
 \begin{split}\label{eq:BDMPS-ININ}
\left(\omega  \frac{dI}{d^2\mathbf{k}d\omega }\right)^{\mathbf{In-In}}&=\frac{2C_F\alpha_s}{(2\pi^2)\omega^{2}(N_c^2-1)^2}\Re\Bigg[\int_{\textbf{x}_g\textbf{x}_g^\prime x^+ x^{+}_c \textbf{z}}e^{i\textbf{k}(\textbf{x}^\prime_g-\textbf{x}_g)}\partial_{\textbf{x}}\mid_0\cdot \partial_{\textbf{x}^\prime}\mid_0 
\\ &\times\Tr \langle G(\textbf{x},\textbf{z}\mid \omega) W(\textbf{0}))
\rangle_{x^+_c,x^+}   \Tr\langle G(\textbf{z},\textbf{x}_g\mid \omega)G^{\dagger}(\textbf{x}^\prime,\textbf{x}^\prime_g \mid\omega) \rangle_{L,x_c^+}\Bigg] \, .
\end{split}
\end{equation}
In the above equation $\omega$ is the frequency of the in-medium gluon, $\omega^\prime$ the energy of the vacuum gluon (the same convention is followed for the transverse momentum $\textbf{k}$). The first term on the right hand side of \eqref{eq:main_amplitude_2} corresponds to the vacuum emission spectrum of a soft gluon. $W$ denotes the adjoint Wilson line over the $+$ direction and $G$ corresponds to the full the gluon propagator including the typical subeikonal corrections taken into account in jet quenchig.\par 
Via direct inspection of the above equations we see that, as expected, the major modification to the spectrum comes in the form of a novel three point functions for the late time evolution of the system. In the harmonic approximation, the tree point function gives rise to the following path integral 
\begin{equation}\label{eq:path_int}
\int_\textbf{z}^{\textbf{x}_g} \mathcal{D}\textbf{r}_1 \int_{\textbf{x}^\prime}^{\textbf{x}_g^\prime}\mathcal{D}\textbf{r}_2 \exp \left(\int_t\frac{i\omega}{2} (\dot{\textbf{r}}_1^2-\dot{\textbf{r}}_2^2) -\frac{\hat{q}}{4 }\int_t\left(\textbf{r}_1\cdot\textbf{r}_2+(\textbf{r}_2-\textbf{r}_1)^2\right)\right) \, ,    
\end{equation}
where $\textbf{r}_1$ denotes the transverse position of the gluon in the amplitude and $\textbf{r}_2$ the position of the gluon in the complex conjugate amplitude. We notice that the term proportional to $(\textbf{r}_2-\textbf{r}_1)^2$ is the same one would obtain in the BDMPS-Z calculation and therefore only the term proportional to the dot product should encapsulate color coherence effects. In order to extract this contribution we expand the gluon propagator in the so called titled Wilson line approximation $G\sim G_0(\textbf{x}_{\rm cl}) W(\textbf{x}_{\rm cl})$, where $G_0$ is the vacuum gluon propagator and $\textbf{x}_{\rm cl}$ is the classical path over which we evaluate the Wilson line. In this approximation, both gluons propagate in a straight line (in transverse space) and we parametrize their trajectories the following way\footnote{Notice that the in the case where the potential in the path integral exponent only has the dipole term, this approximation does not modify the result.} ($\tau=x_c^+-x^+$)
\begin{equation}
\textbf{r}_1(t)\sim \textbf{x}+(t+\tau)\theta   \qquad
\textbf{r}_2(t)\sim\textbf{x}^\prime+t\theta \qquad\Rightarrow   
\textbf{r}_1(\Delta t)\cdot \textbf{r}_2(\Delta t)\sim (\Delta t)^2 \theta^2 \, .
\end{equation}
In the above equations $\theta$ is the emission angle and we have dropped terms that scale with the formation time, which is much smaller than $\Delta t=L-x_c^+$, the typical time over which the particle broadens. Combining this parametrization with the tilted Wilson line approximation and \eqref{eq:path_int} we obtain that the spectrum \eqref{eq:main_amplitude_2}
can be written as the BDMPS-Z result controlled by the following term
\begin{equation}\label{eq:deltamed_sec3}
1-\Delta_{med}\equiv \exp\left(-\frac{\hat{q}}{12}\theta^2(\Delta t)^{3}\right)    \, ,
\end{equation}
which is qualitatively equivalent to \eqref{eq:deltam}. Notice that in obtaining \eqref{eq:deltamed_sec3} we have ignored that during the time $\Delta t$, the outgoing gluon suffers additional broadening during the same time interval. Therefore \eqref{eq:deltamed_sec3} enters the evolution of the shower as a correction to the in-medium splitting kernel. This is in line with the Markovian approximation for multiple gluon emission, where the broadening and emission kernels are effectively decoupled.

\section{Including color coherence into the rate equations}
The rate equations \cite{BDIM2} describe the in-medium shower as is if it consisted of multiple decoherent gluon emissions. The kernel of emission for each gluon and its future momentum broadening in the medium are obtained directly from the BDMPS-Z calculation \cite{BDIM1,Liliana2}. In this approximation the functional that generates the medium induced shower reads \cite{BDIM2}
\begin{equation}\label{eq:Z}
\begin{split}
\partial_t \mathcal{Z}_{p_0}(t,t_0\mid u)&=\int_\textbf{l} C(\textbf{l},t_0) u(p_0^+,\textbf{p}_0+\textbf{l})
\\&+\alpha_s\int_z \mathcal{K}(\textbf{l},z,p_0^+)\left[u(zp_0^+, z\textbf{p}_0+\textbf{l})u((1-z)p_0^+,(1-z)\textbf{p}_0-\textbf{l})-u(p_0^+,\textbf{p}_0-\textbf{l})\right] \, .
\end{split}
\end{equation}
The term entering at order $\mathcal{O}(\alpha_s)$ corresponds to the single particle momentum broadening and is controlled by the (Fourier transform of the) broadening kernel $C(\textbf{r},t_0)=-\frac{\hat{q}}{4}\textbf{r}^2$. The term proportional to $\alpha_s$ describes the splitting of the incoming state into two outgoing partons, via the splitting kernel $\mathcal{K}$. The term with a minus sign ensures unitarity.  \par 
One can rederive \eqref{eq:Z} but including the factor \eqref{eq:deltamed_sec3} and taking into account the diagram where the blue gluon (see figure \ref{fig:1})  couples to the same emitter in amplitude and conjugate amplitude. In the end, one obtains the new evolution equation which has the same form as \eqref{eq:Z} but with a novel splitting kernel 
\begin{equation}\label{eq:ckprime}
\mathcal{K}^\prime(\textbf{l},z,p_0^+,t-t_0)=\overbrace{\frac{2}{p_0^+}\frac{P_{gg}(z)}{z(1-z)}\sin\left(\frac{\textbf{l}^2}{2k_f^2}\right)\exp\left(-\frac{\textbf{l}^2}{2k_f^2}\right)}^{\mathcal{K}(\textbf{l},z,p_0^+)}
\overbrace{\left\{1-\exp\left(-\frac{\hat{q}}{12}\theta^2(z,\textbf{l},p_0^+)(t-t_0)^3\right)\right\} }^{\Delta_{med}(z,\textbf{l},p_0^+,t-t_0)} \, .
\end{equation}
 Here we define the emission angle as $\theta(\textbf{l},z)=\frac{\textbf{l}}{(1-z)z}$ and $k_f^2=\sqrt{\hat{q}(1-z(1-z))p_0^+(z(1-z))}$ is the typical transverse momentum acquired during the branching process. Notice that the physical picture detailed in the previous sections is in fact enforced by this kernel of emission: if the emission angle is large then $\mathcal{K}^\prime\to\mathcal{K}$, while for not so large angles, the new term diminishes the probability of emission. In the limit where $\theta\to0$ (out of the domain of applicability of the rate equations) $\mathcal{K}^\prime\to 0$, i.e. a new emission is forbidden since the new 2-particle system would behave has if they were a single color charge. Also notice that (as advertised above)
\begin{equation}
\int_{0}^L \mathcal{K}^\prime(t)\sim L-t_d+\mathcal{O}\left(L\frac{t_d^3}{L^3}\right)  \, .
\end{equation}

\section{Conclusion and Outlook}
We have presented a simple (phenomenology oriented) way of introducing color coherence effects into the rate equations which govern multiple soft gluon in-medium branching. The result obtained obeys the expected scalings and it is of easy physical interpretation. \par 
In the future, it would be interesting to perform a numerical study implementing \eqref{eq:ckprime} and explore its impact on observables sensitive to the angular distribution of radiation.
\acknowledgments{
The authors are grateful to Néstor Armesto for helpful discussions. The project that gave rise to these results received the support of a fellowship from ``la Caixa" Foundation (ID 100010434). The fellowship code is LCF/BQ/ DI18/11660057. This project has received funding from the European Union's Horizon 2020 research and innovation programm under the Marie Sklodowska-Curie grant agreement No. 713673. We were supported by Ministerio de Ciencia e Innovacion of Spain under project FPA2017-83814-P; Unidad de Excelencia Maria de Maetzu under project MDM-2016-0692; European research Council project ERC-2018-ADG-835105 YoctoLHC; and Xunta de Galicia (Conselleria de Educacion) and FEDER.}

\bibliographystyle{JHEP}
\bibliography{Lib.bib}

\end{document}